\documentclass[a4paper,10pt,twoside]{cpc-hepnp}

\usepackage{multicol}
\usepackage{graphicx}
\usepackage{booktabs}
\usepackage{amssymb,bm,mathrsfs,bbm,amscd}
\usepackage[tbtags]{amsmath}
\usepackage{lastpage}

\usepackage[english]{babel}

\newcommand{\ie}{\emph{i}.\emph{e}.}
\newcommand{\Pt}{$p_{t}$}

\usepackage{times}
\usepackage{scrtime}

\usepackage{array}
\newcommand{\PreserveBackslash}[1]{\let\temp=\\#1\let\\=\temp}
\newcolumntype{C}[1]{>{\PreserveBackslash\centering}p{#1}}
\newcolumntype{R}[1]{>{\PreserveBackslash\raggedleft}p{#1}}
\newcolumntype{L}[1]{>{\PreserveBackslash\raggedright}p{#1}}

\usepackage{xcolor}
\begin{document}

\fancyhead[co]{\footnotesize Ma Changli~ et al: An extended segment
  pattern dictionary for pattern matching tracking algorithm at BESIII}

  \footnotetext[0]{Received \today}

  \title{An extended segment pattern dictionary for pattern matching
    tracking algorithm at BESIII\thanks{
      Supported by
	Ministry of Science and Technology of China (2009CB825200), Joint
	Funds of National Natural Science Foundation of China
	(11079008, 11121092), Natural Science Foundation of China (10905091)
	and SRF for ROCS of SEM.
    }
  }

\author{%
  MA~Changli$^{1}$%
    \quad  ZHANG~Yao$^{2}$
    \quad  YUAN~Ye$^{2;1)}$\email{yuany@ihep.ac.cn}%
    \quad  LU~Xiao-Rui$^{1;2)}$\email{xiaorui@gucas.ac.cn}%
    \\ZHENG~Yangheng$^{1;3)}$\email{zhengyh@gucas.ac.cn}%
    \quad  He~Kangli$^{2}$
    \quad  Li~Weidong$^{2}$
    \\Liu~Huaiming$^{2}$
    \quad  MA~Qiumei$^{2}$
    \quad  Wu~Linghui$^{2}$
}

\maketitle

\address{%
  $^{1}$ Graduate University of Chinese Academy of Sciences, Beijing 100049, P. R. China\\
    $^{2}$ Institute of High Energy Physics, Beijing 100049, P. R. China\\
}

\begin{abstract}
A pattern matching based tracking algorithm, named MdcPatRec, is
used for the reconstruction of charged tracks in the drift chamber
of the BESIII detector.  This paper addresses the shortage of
segment finding in MdcPatRec algorithm. An extended segment
construction scheme and the corresponding pattern dictionary are
presented. Evaluation with Monte-Carlo and experimental data show
that the new method can achieve higher efficiency for low transverse
momentum tracks.
\end{abstract}



\begin{multicols}{2}

\section{Introduction}
The Beijing Electron Positron Collider II (BEPCII)~\cite{bes3} is a
double-ring multi-bunch $e^{+}e^{-}$ collider which is operated in
the collision energy range from 2\,GeV to $4.6$\,GeV. The Beijing
Spectrometer III (BESIII)~\cite{bes3} is a general-purpose detector
with $93\%$ coverage of full
solid angle. BEPCII and BESIII are powerful facilities for the study
of charmonium physics, D-physics, hadron spectroscopy and $\tau$
physics~\cite{physics}. From the interaction area of the $e^{+}e^{-}$
beams to the outside, there are 5 apparatuses: a drift chamber
(MDC)~\cite{design_report}, which includes 43 layers of drift cells
and is used for charged tracking; a time-of-flight counter, which
is composed of 2 layers of scintillator bars for measuring the
flight time of penetrating charged particles; an electromagnetic
calorimeter, which comprises 6240 CsI(Tl) crystals for the measurement
of photon energy; a super-conducting solenoid magnet, which provides
a 1 Tesla magnetic field and a muon counter, which includes 1000\,m$^{2}$
resistive plate chambers, used for identifying muon.

MDC is a carbon fiber cylinder flowed with a mixture of He and
C$_{3}$H$_{8}$ (He/C$_{3}$H$_{8}$ = 60/40) as working gas, and with
signal and field wires strung between the two end-caps.
The signal wires are operated with a high positive voltage, while
the field wires are at ground voltage.  There are 6796 signal wires
in total; each of them is surrounded by 8 or 9 field wires.  The
signal wires are grouped into 43 layers, and these layers are
divided into 11 super-layers, with the labels SL-$J$ (from the
    inside to the outside, $J=$ 1, 2, $\cdots$, 11.). The SL-11
consists of only 3 signal layers, while each of the other 10
super-layers includes 4 signal layers.  In some super-layers, the signal
wires are parallel to the $z$ axis of the MDC, so these super-layers are
named axial super-layers. In the other super-layers, the signal wires have a
stereo angle with respect to the $z$ axis of the MDC, and these
super-layers are named stereo super-layers.  Axial (A) and stereo
(U V, U means the stereo angle is negative, V means the stereo angle
 is positive) super-layers follow the order
UVAAAUVUVAA~\cite{design_report}.  The small space around a signal
wire and surrounded by the neighbouring field wires is called a
drift cell.  The sections of these drift cells are nearly square
in the plane perpendicular to the beam axis, with the size of
approximately $12\times12$\,mm$^{2}$ in the innermost two
super-layers and $16\times16$\,mm$^{2}$  in the other super-layers.
The signal wires collect the avalanche electrons produced by
ionization along the particle trajectories. Electric signals on fired
wires are called hits and are used to reconstruct trajectories
of charged particles.

One basic offline \texttt{C++} software algorithm for MDC track
reconstruction at BESIII is a pattern matching based tracking package
(MdcPatRec)~\cite{Ref_PAT}, which extracts track segments from a
set of hits within a group of neighboring cells and links the
segments into tracks.  For charged particles with high transverse
momentum (\Pt), the tracking efficiency of MdcPatRec is satisfied
for physics analysis.  However, in low \Pt~ range, the efficiency
still has a large space to improve.  In this paper we propose an
extended segment construction scheme and its pattern dictionary
which is justified to be helpful in the improvement of low
\Pt~ tracking efficiency.

\section{MdcPatRec algorithm}

To introduce the procedure of MdcPatRec, we list the important
glossaries first:

(1) Segment.
In a super-layer, a series of hits of a charged particle are
defined as a segment. The arc in Fig. \ref{figure_mdc_segment_8w}
shows a track.  According to the segment definition, the 4 hits at
the 4 fired wires compose a segment.

\begin{center}
\includegraphics[height=0.40\linewidth,width=0.95\linewidth]
{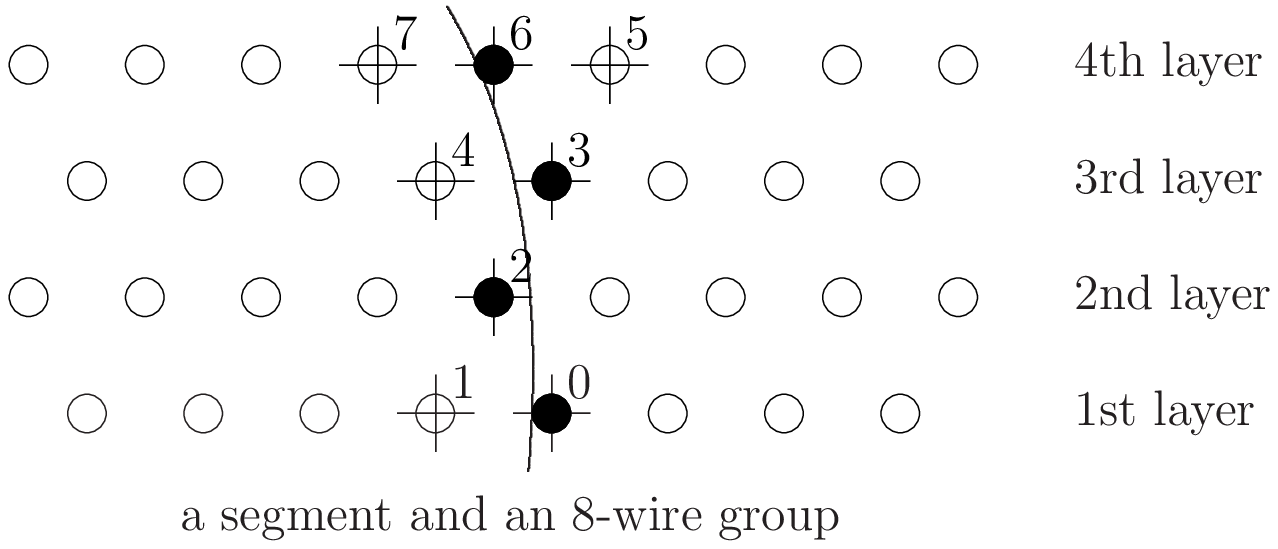}
\renewcommand{\figurename}{Fig.}
\figcaption
{
  \label{figure_mdc_segment_8w}%
    A schematic plot for illustrating a segment and an 8-wire group.
    The 4 layers of circles denote the signal wires in one super-layer.
    The arc shows a track, and the black solid circles are the fired
    wires of the particle. The 8 adjacent signal wires marked with
    a cross compose an 8-wire group used to match the segment pattern
    in segment finding.
}
\end{center}

(2) Segment pattern and segment pattern dictionary.
The fired wire distribution of a segment in a super-layer is defined
as a segment pattern. In present MdcPatRec, segment finder uses the hits
in the 2nd layer of a super-layer as seeds to search
for segments in every group of 8 signal wires, the relative position
of these signal wires is graphically represented by the 8 wires
marked by a cross in Fig. \ref{figure_mdc_segment_8w}. A segment pattern
is expressed by 8 bits of an integer in \texttt{C++} program: the fired wires
are indicated by ``1'', while the non-fired wires are indicated by ``0''.
For instance, the segment pattern illustrated in Fig. \ref{figure_mdc_segment_8w}
is expressed with 01001101 in the binary number (77 in the decimal number).
Segment patterns are collected to build a segment pattern dictionary.

The procedure of MdcPatRec includes the following steps: first to find
segments in all super-layers by pattern matching; then to assemble segments
in axial super-layers into circular tracks and apply a circle fit.
Second, to add segments in stereo super-layers to circle tracks to
constitute helix tracks and apply helix fit. After helix fit, tracks
are reconstructed, and they are stored for Kalman fit, tracks
extrapolation, physics analysis and so on.

\section{The shortage of 8-wire segment pattern in MdcPatRec}\label{shortage_dictrionary}

The tracking efficiency of MdcPatRec is satisfied for non-curled
tracks in the drift chamber.  It loses some efficiency for low
\Pt~ tracks due to the reduced efficiency for segment finding
within a super-layer. Azimuth coverage angle of segment groups
does not meet the requirements for low \Pt~ track segment finding.

Drift cell size limits the azimuth coverage of segment pattern
dictionary. We compare the azimuth coverage of an 8-wire segment
group with the azimuth coverage of tracks in a super-layer to
test whether the segment group is large enough to reconstruct
all possible segments.  We define a variable $\Delta{\phi}$ to
describe the coverage of a segment group along $\phi$
  direction, the definition of $\Delta{\phi}$ is described by
Eq. (\ref{eq_hit_span_seg})
  \begin{eqnarray}
  \label{eq_hit_span_seg}
  \Delta{\phi}&=& |\phi_{L-L(L-R)} - \phi_{U-R(U-L)}|, %
  \end{eqnarray}
  \noindent where $\phi_{L-L}$ is the $\phi$ angle of the drift cell
  at the low left corner of an 8-wire group, as displayed in Fig.
  \ref{figure_mdc_segment_8w}; similarly, $L-R$ means the low right
  corner, $U-L$ means the upper left corner, and $U-R$ means the
  upper right corner.  The $\Delta{\phi}$ values of 8-wire segment
  pattern dictionary are listed in Table \ref{table_hit_span_seg}.
  We also define a variable, $\delta{\phi}$, to describe the
  azimuth coverage of a track in every super-layers along $\phi$
  direction. In the magnetic field, $\delta{\phi}$ of a track with
  certain \Pt~ within a super-layer can be calculated by Eq.
  (\ref{eq_hit_span_pt}).
  \begin{eqnarray}
  \label{eq_hit_span_pt}
  \delta{\phi}&=&
  | \arctan( \dfrac{r_{in}^{2}}{\sqrt{4r_{p}^{2}r_{in}^{2} - r_{in}^{4}}} ) - \nonumber \\[1mm]
  &&
  \arctan( \dfrac{r_{out}^{2}}{\sqrt{4r_{p}^{2}r_{out}^{2} - r_{out}^{4}}} ) |.
  \end{eqnarray}
  \noindent In  Eq. (\ref{eq_hit_span_pt}), $r_{in}$ and $r_{out}$ are the
  inner radius and outer radius of a super-layer respectively; $r_{p}$ is
  the radius of a track with certain
\Pt, which can be calculated by Eq. (\ref{eq_radius_pt})
  \begin{eqnarray}
  \label{eq_radius_pt}
  r_{p}&=& \dfrac{p_{t}}{B_{z}\cdot q}, %
  \end{eqnarray}
  \noindent where $B_{z}$ is the magnetic field intensity along the beam
  axis and $q$ is the electric charge of the particle; all variables are
  expressed in the international system of units. The calculated
  $\delta{\phi}$ are listed in Table \ref{table_hit_span_pt}. Comparing
  these $\delta{\phi}$ with $\Delta{\phi}$ listed in Table
  \ref{table_hit_span_seg}, we find that the 8-wire segment pattern dictionary
  is not large enough to cover all segments of some low \Pt~ charged particles.

The wire groups used to search for segments in the MDC are not all
symmetric along $\phi$ direction. If the amounts of wires of two
adjacent layers are the same, the two layers are staggered
exactly by a half drift cell along $\phi$ direction, as shown
in Fig. \ref{figure_mdc_segment_8w}. Otherwise, the two layers
cannot be placed in the exact staggered form. As a result, the
symmetric arrangement of the wire groups in a super-layer is
determined by the amounts of wires in the four layers. From SL-1
to SL-5, the amounts of wires in each layer are not the same, and
some 8-wire wire groups in these super-layers are asymmetric.
Figure \ref{figure_segment_form_real} shows two asymmetric 8-wire
wire groups in SL-1.  As shown in these plots, the azimuth
coverage of an asymmetric  8-wire wire group is smaller than
that of a symmetric one.  Table \ref{table_hit_span_seg} lists
the minimum wire group coverages of 8-wire segment pattern
dictionary in every super-layer; as expected, the asymmetry of
wire groups in SL-1 to SL-5 reduces the coverage of segment pattern.
In addition, the 8-wire segment pattern dictionary is designed
based on the hypothesis of symmetric wire groups, and the
dictionary doesn't suit the asymmetric wire groups.

\begin{center}
\includegraphics [height=0.12\textheight,width=0.49\linewidth]
{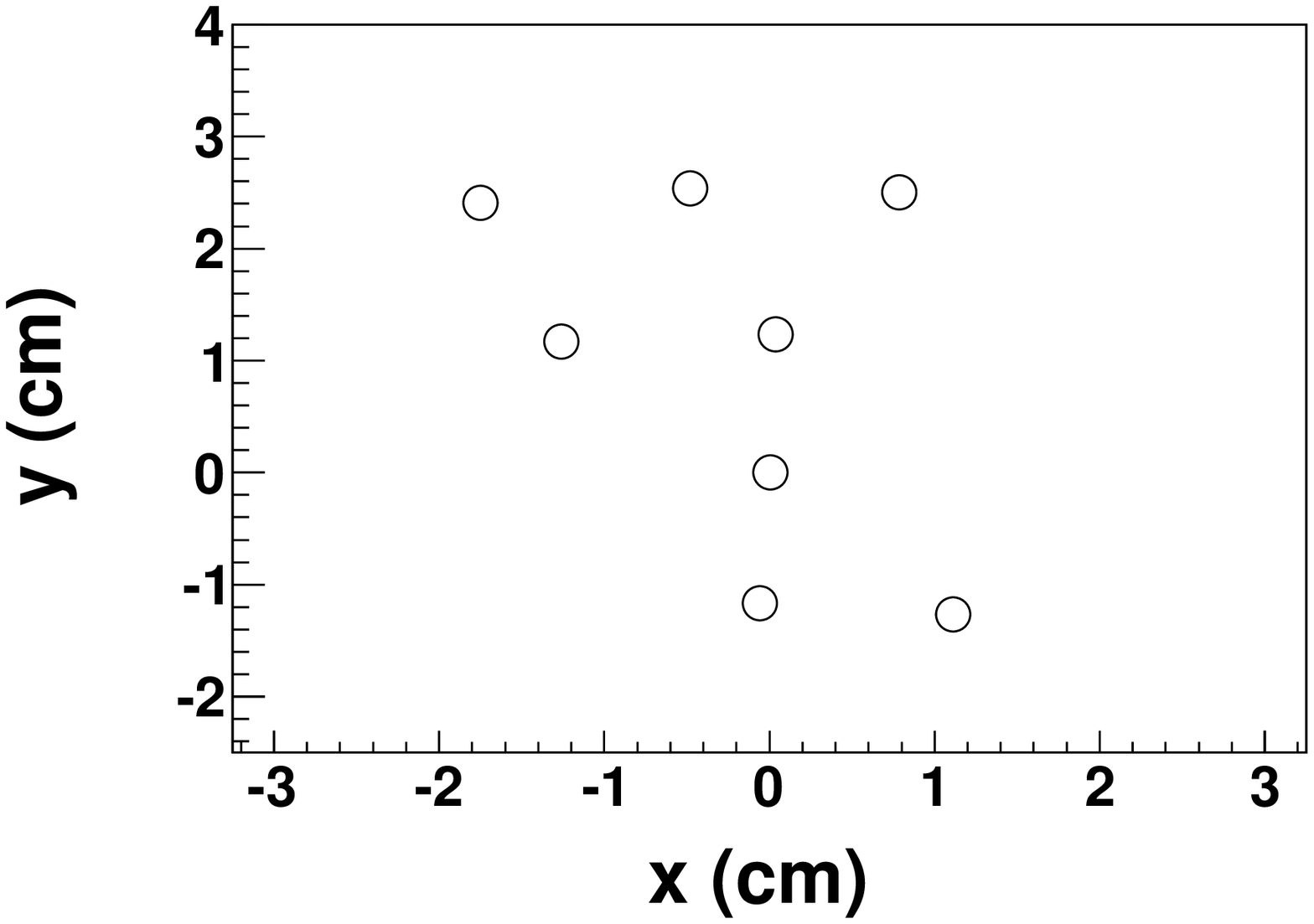}
\includegraphics [height=0.12\textheight,width=0.49\linewidth]
{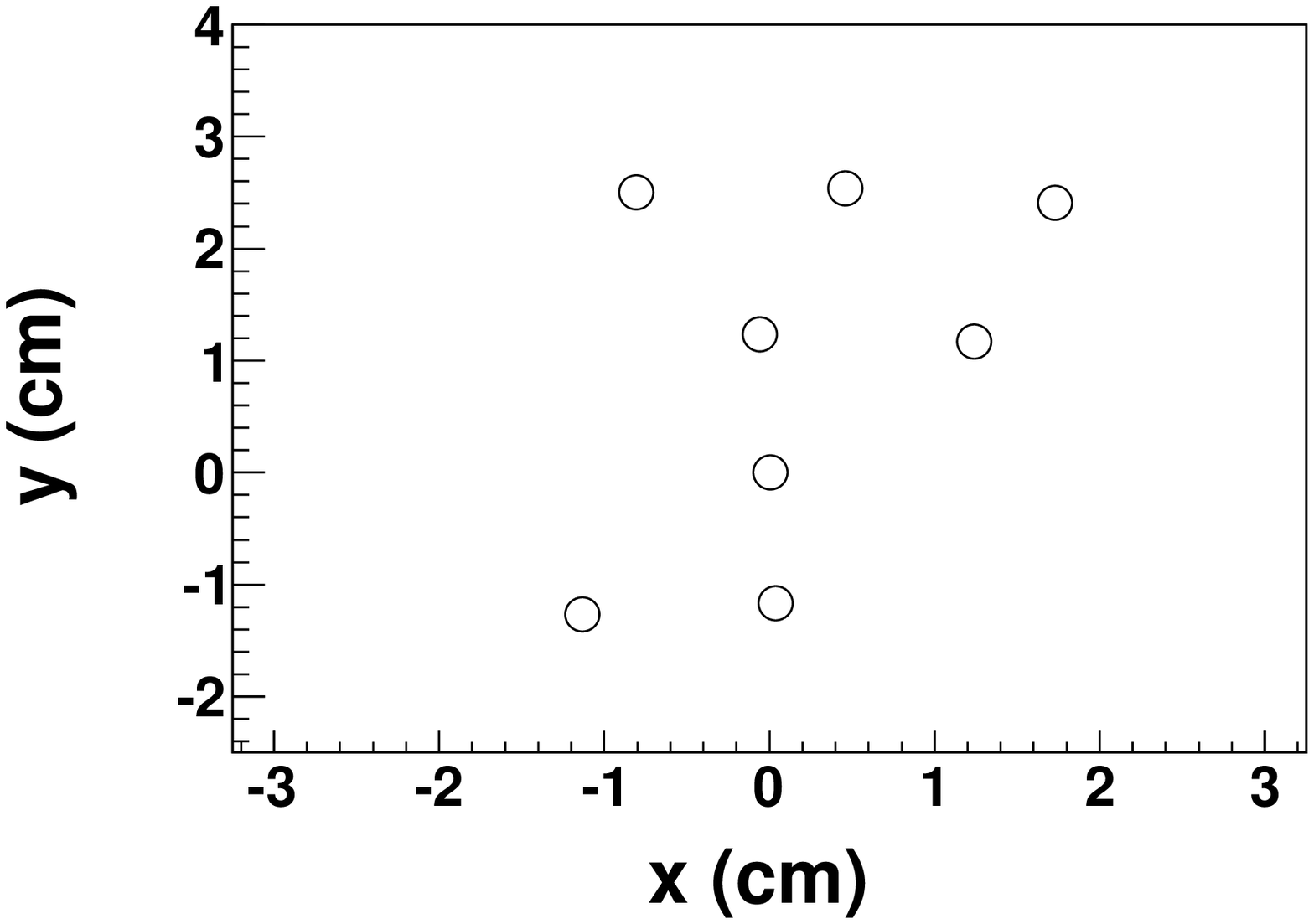}
\renewcommand{\figurename}{Fig.}
\figcaption{
  \label{figure_segment_form_real}
  The relative position of signal wires of 2 asymmetric 8-wire groups
    in SL-1. The origin of coordinate in each plot is set at the
    position of the signal wire in the second layer.
}
\end{center}
\ruleup
  \end{multicols}
  \begin{center}
  \tabcaption{
    \label{table_hit_span_seg}
    The azimuth coverages, \ie \ $\Delta{\phi}$ (degree), of 8-wire and 14-wire
      segment pattern dictionary in each super-layer. The `min' value is the
      minimum coverage in a super-layer. The wire groups are not all symmetric
      in SL-1 to SL-5, so the minimum $\Delta{\phi}$ is smaller than the symmetric
      wire groups. In SL-6 to SL-11, all wire groups are symmetric, so the
      minimum $\Delta{\phi}$ is equal to that of the symmetric wire groups.
  }
{\footnotesize
  \begin{tabular}{c C{1.6eM}  C{1.6eM}   C{1.6eM}   C{1.6eM}   C{1.6eM}C{1.6eM}  C{1.6eM}   C{1.6eM}   C{1.6eM}   C{1.6eM}C{1.6eM}}
\toprule
\multicolumn{1}{c}{Dictionary} &
\multicolumn{1}{c}{SL-1} &
\multicolumn{1}{c}{SL-2} &
\multicolumn{1}{c}{SL-3} &
\multicolumn{1}{c}{SL-4} &
\multicolumn{1}{c}{SL-5} &
\multicolumn{1}{c}{SL-6} &
\multicolumn{1}{c}{SL-7} &
\multicolumn{1}{c}{SL-8} &
\multicolumn{1}{c}{SL-9} &
\multicolumn{1}{c}{SL-10} &
\multicolumn{1}{c}{SL-11}
\\
\hline
 8-wire & 18.6 & 12.4 & 10.9 & 8.4 & 6.7 & 5.6 & 5.1 & 4.3 & 3.8 & 3.5 & 2.5
\\
 8-wire   min & 10.9 & 7.9 & 6.8 & 5.2 & 4.1 &  ---  &  ---  &  ---  &  ---  &  ---  &  ---
\\
\hline
 14-wire & 34.1 & 22.5 & 19.7 & 15.2 & 12.1 & 10.1 & 9.2 & 7.8 & 6.8 & 6.3 & 5.0
\\
 14-wire  min & 26.4 & 18.0 & 15.6 & 12.0 & 9.5 &  ---  &  ---  &  ---  &  ---  &  ---  &  ---
\\
\bottomrule
\end{tabular}

}
\end{center}

\begin{center}
\tabcaption{
  \label{table_hit_span_pt}
  The track azimuth coverage, \ie \ $\delta{\phi}$ (degree), of charged particles
    with different \Pt~ in every super-layer.  If the \Pt~ of a particle is
    smaller than 120\,MeV/c, the particle curls in the MDC. If the $\delta{\phi}$
    is larger than the minimum coverage of 8-wire segment pattern dictionary
    in this super-layer, it is marked with an underline.  If the $\delta{\phi}$
    is larger than the minimum coverage of 14-wire segment pattern dictionary
    in this super-layer, it is marked with an overline.
}
{\footnotesize
  \begin{tabular}{c C{1.6eM}  C{1.6eM}   C{1.6eM}   C{1.6eM}   C{1.6eM}C{1.6eM}  C{1.6eM}   C{1.6eM}   C{1.6eM}   C{1.6eM}C{1.6eM}}
\toprule 
\multicolumn{1}{c}{$p_{t}$ (MeV/c)} &
\multicolumn{1}{c}{SL-1} &
\multicolumn{1}{c}{SL-2} &
\multicolumn{1}{c}{SL-3} &
\multicolumn{1}{c}{SL-4} &
\multicolumn{1}{c}{SL-5} &
\multicolumn{1}{c}{SL-6} &
\multicolumn{1}{c}{SL-7} &
\multicolumn{1}{c}{SL-8} &
\multicolumn{1}{c}{SL-9} &
\multicolumn{1}{c}{SL-10} &
\multicolumn{1}{c}{SL-11} 
\\
\hline
50 & 6.7 & 6.6 & \underline{11.2} & $\overline{16.9}$ &     &     &     &     &     &     &    
\\
70 & 4.7 & 4.5 & 6.8 & \underline{7.5} & \underline{8.9} & $\overline{14.0}$ &     &     &     &     &    
\\
90 & 3.6 & 3.4 & 5.0 & \underline{5.3} & \underline{5.6} & \underline{6.4} & \underline{8.0} & $\overline{11.9}$ &     &     &    
\\
110 & 2.9 & 2.8 & 4.0 & 4.1 & \underline{4.2} & 4.5 & \underline{5.1} & \underline{5.7} & \underline{6.7} & $\overline{12.0}$ &    
\\
130 & 2.5 & 2.3 & 3.3 & 3.4 & 3.4 & 3.6 & 3.9 & 4.1 & \underline{4.4} & \underline{5.4} & \underline{4.1}
\\
150 & 2.1 & 2.0 & 2.9 & 2.9 & 2.9 & 3.0 & 3.2 & 3.3 & 3.4 & \underline{3.9} & 2.7
\\
170 & 1.9 & 1.8 & 2.5 & 2.5 & 2.5 & 2.6 & 2.8 & 2.8 & 2.8 & 3.1 & 2.1
\\
\bottomrule
\end{tabular}

}
\end{center}
\vspace{0.5cm}

\ruledown \vspace{0.5cm}

\begin{multicols}{2}

\section{The extended segment pattern dictionary}
In Sec.\,3, we explain 8-wire segment pattern dictionary
which is
not optimal in the azimuth coverage for tracks with low \Pt.
In addition, the asymmetry of the wire groups aggravates this
problem. Hence, it is meaningful to broaden the 8-wire groups to
larger wire groups and design an extended segment pattern
dictionary for segment finding.

An easy way to broaden the coverage of the 8-wire segment
pattern dictionary is to add a signal wire at each side of
an 8-wire group on the 1st, the 3rd and the 4th layer, and an 8-wire
group is extended to a 14-wire group as illustrated in
Fig. \ref{figure_mdc_segment_14w}. Comparing the azimuth
coverage of particle tracks in Table \ref{table_hit_span_pt}
with the coverage of 14-wire segment pattern dictionary in
Table \ref{table_hit_span_seg}, we can find that the 14-wire
segment pattern dictionary can cover all segments of different
tracks except the segments in the outermost super-layer, which
a track can reach. This type of segments has very large
azimuth coverage angle and cannot be reconstructed in an easy way.

A new segment pattern dictionary is built based on the 14-wire
groups. Because the least squares fitting is applied to the hits
of a segment pattern in segment finding, a segment pattern
requires 3 or more hits. Therefore we build the dictionary
with 3-hit and 4-hit segment patterns. A 4-hit segment pattern
has a hit in every layer, and a 3-hit pattern has 3 hits in
three different layers. For the 4-hit segment patterns, there are
80 ($4 \times 1 \times 4 \times 5$) combinations, but some
combinations are obviously impossible and we remove them, which
are determined by the relative position of the signal wires in a
14-wire group. From SL-6 to SL-11, all wire groups are
symmetric, and we build a unique segment pattern dictionary for
them.  There is no significant asymmetry from SL-2 to SL-5 and
we find that the dictionary based on symmetric wire groups is
applicable. In SL-1, the asymmetry of wire groups is more
serious than any other super-layers, and we build a special
segment pattern dictionary. Finally the segment pattern
dictionary for super-layers from SL-2 to
SL-11 includes 30 4-hit segment patterns and 56 3-hit segment
patterns, and the segment pattern dictionary for SL-1 includes
44 4-hit segment  patterns and 72 3-hit segment  patterns.
Fig. \ref{plot_seg_dict_14w_4hit_pattern_0}
shows 6 4-hit segment patterns. In \texttt{C++} program, a
14-wire segment pattern is expressed by 14 bits in an integer
variable: the fired wires are marked by ``1'', while the other
signal wires are marked by ``0''. Table \ref{table_dict_pat_14w_4hit_5layer}
lists the values of decimal numbers corresponding to the segment
patterns drawn in Fig. \ref{plot_seg_dict_14w_4hit_pattern_0}.

\begin{center}
\includegraphics[height=0.40\linewidth,width=0.95\linewidth]
{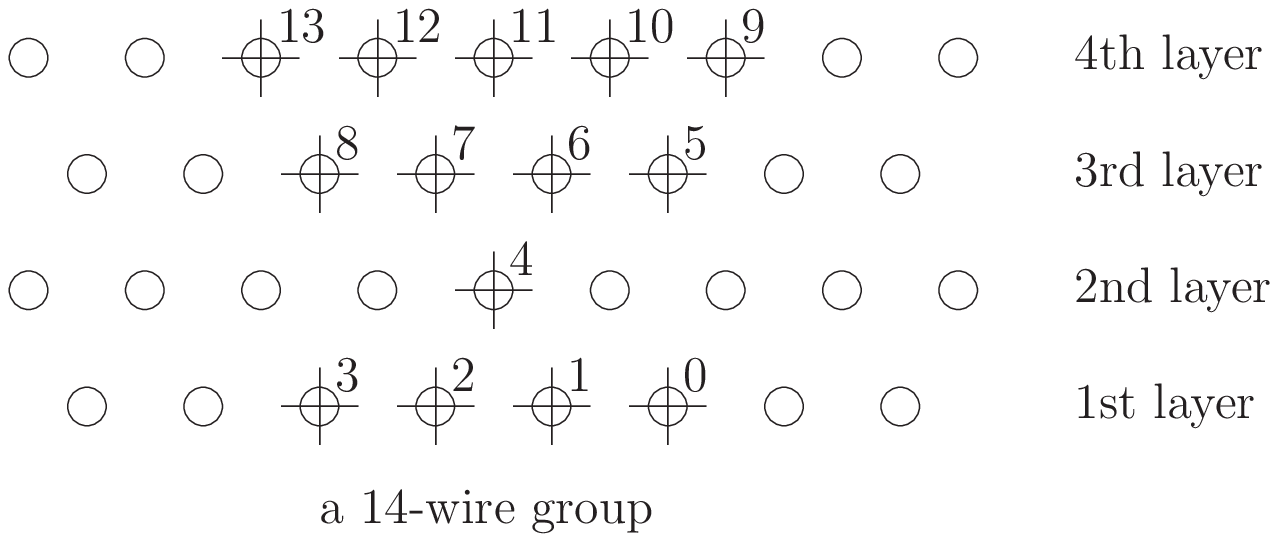}
\renewcommand{\figurename}{Fig.}
\figcaption{
  \label{figure_mdc_segment_14w}
  A 14-wire group schematic plot. The 4 layers of circles denote
    the signal wires in a super-layer. The 14 adjacent signal wires
    marked with a cross make up the 14-wire group.
}
\end{center}

\begin{center}
\includegraphics [height=0.22\textheight,width=0.80\linewidth]
{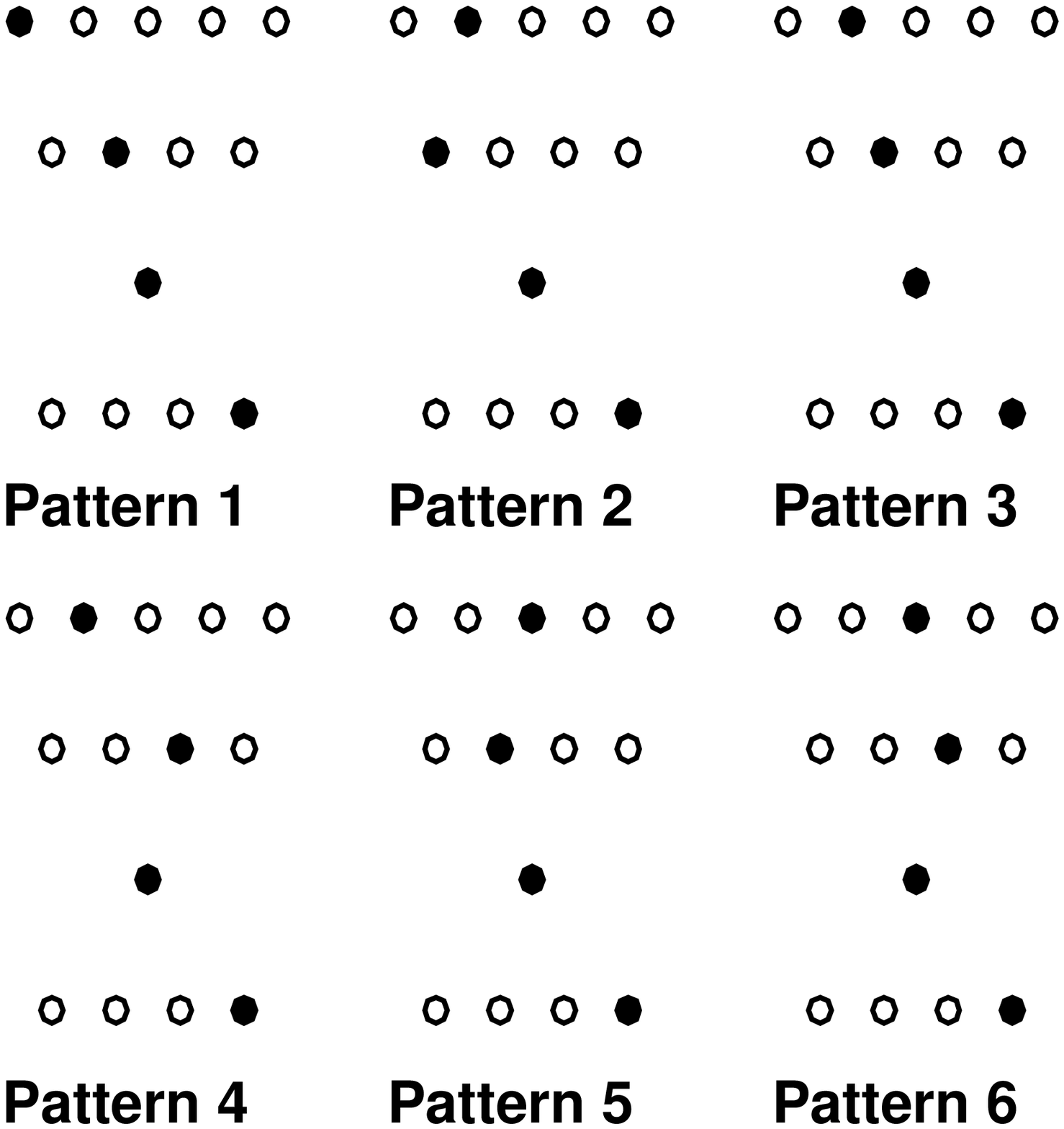}
\renewcommand{\figurename}{Fig.}
\figcaption{
  \label{plot_seg_dict_14w_4hit_pattern_0}
  Some 4-hit 14-wire segment  patterns.
}
\end{center}

\ruleup
\end{multicols}
\begin{center}
\tabcaption{
  \label{table_dict_pat_14w_4hit_5layer}
  The expression of the 4-hit segment patterns listed in
    Fig. \ref{plot_seg_dict_14w_4hit_pattern_0}.
}
{\footnotesize

\begin{tabular}{c c c| c c c | c c c }
\toprule 
\multicolumn{1}{c}{pattern} & 
\multicolumn{1}{c}{decimal} & 
\multicolumn{1}{c|}{binary } & 
\multicolumn{1}{c}{pattern} & 
\multicolumn{1}{c}{decimal} & 
\multicolumn{1}{c|}{binary } & 
\multicolumn{1}{c}{pattern} & 
\multicolumn{1}{c}{decimal} & 
\multicolumn{1}{c}{binary }  
\\
\hline
 1&   8337  &   10000010010001 &  
 2&   4369  &   01000100010001 &   
 3&   4241  &   01000010010001  \\  
 4&   4177  &   01000001010001 &   
 5&   2193  &   00100010010001 &  
 6&   2129  &   00100001010001  \\ 
\bottomrule
\end{tabular}

}
\end{center}
\vspace{0.5cm}

\ruledown \vspace{0.5cm}
\begin{multicols}{2}

\section{The 14-wire segment pattern dictionary performance}
With single charged particle Monte-Carlo (MC) data sample, a multi-prong
physics MC data sample and experimental data sample, the performance of
14-wire segment pattern dictionary is tested.

\subsection{The segment efficiency}
The segment efficiency $\varepsilon_{\mathrm{seg}}$ in a super-layer is
defined as Eq. (\ref{eq_eff_seg}) %
\begin{eqnarray}
\label{eq_eff_seg}
\varepsilon_{\mathrm{seg}}  &=&
\dfrac{N^{\mathrm{rec}}_\mathrm{seg}}{ N^{\mathrm{MC}}_\mathrm{seg}}, %
\end{eqnarray}
\noindent where $N^{\mathrm{rec}}_\mathrm{seg}$ is the number of
segments that have been found in a super-layer, and
$N^{\mathrm{MC}}_\mathrm{seg}$ is the number of tracks passing through
the super-layer in MC data sample. Figure \ref{eff_eff_seg_Psi3770_single_e_neg}
shows the segment efficiency of $e^{-}$ with different \Pt, as shown
in these plots, the 14-wire segment pattern dictionary can achieve higher
segment efficiency at the super-layer which track curled back.

\subsection{The tracking efficiency}
We generate 400,000 single $e^{-}$ MC data sample to test the improvement
of tracking efficiency. In the MC data sample, the momentum of $e^{-}$ has
a uniform distribution from 0.05 to 1\,GeV/c, and the angular distribution of
the $e^{-}$  is uniform in full solid angle.
Tracking efficiency
$\varepsilon_{\mathrm{trk}}$ is defined as Eq. (\ref{eq_eff_trk}) %
\begin{eqnarray}
\label{eq_eff_trk}
\varepsilon_{\mathrm{trk}}  &=& \dfrac{N^{\mathrm{rec}}_\mathrm{trk}}{ N^{\mathrm{MC}}_\mathrm{trk}}, %
\end{eqnarray}
\noindent where $N^{\mathrm{rec}}_\mathrm{trk}$ is the number of
events that have reconstructed a charged track, and
$N^{\mathrm{MC}}_\mathrm{trk}$ is the number of events that have a
charged track in MC data sample. Figure \ref{graph_eff_trk_ZeroVSSix_e_neg_trk_rec_range01}
shows the tracking efficiency for $e^{-}$. As shown in these
plots, the tracking efficiency of 14-wire segment pattern dictionary
is higher than that of the 8-wire segment pattern dictionary.
\begin{center}
\includegraphics [height=0.16\textheight,width=0.49\linewidth]
{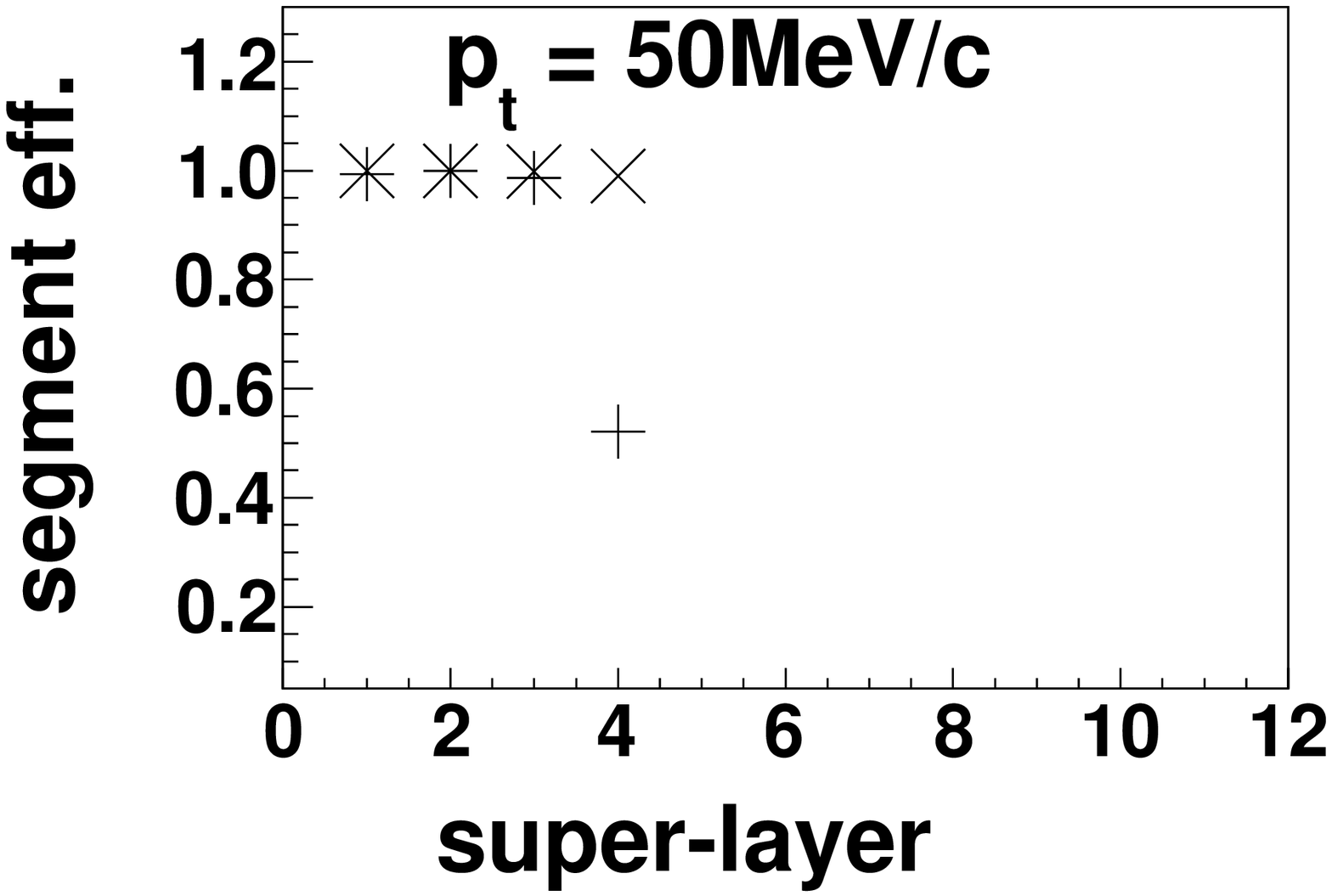}
\includegraphics [height=0.16\textheight,width=0.49\linewidth]
{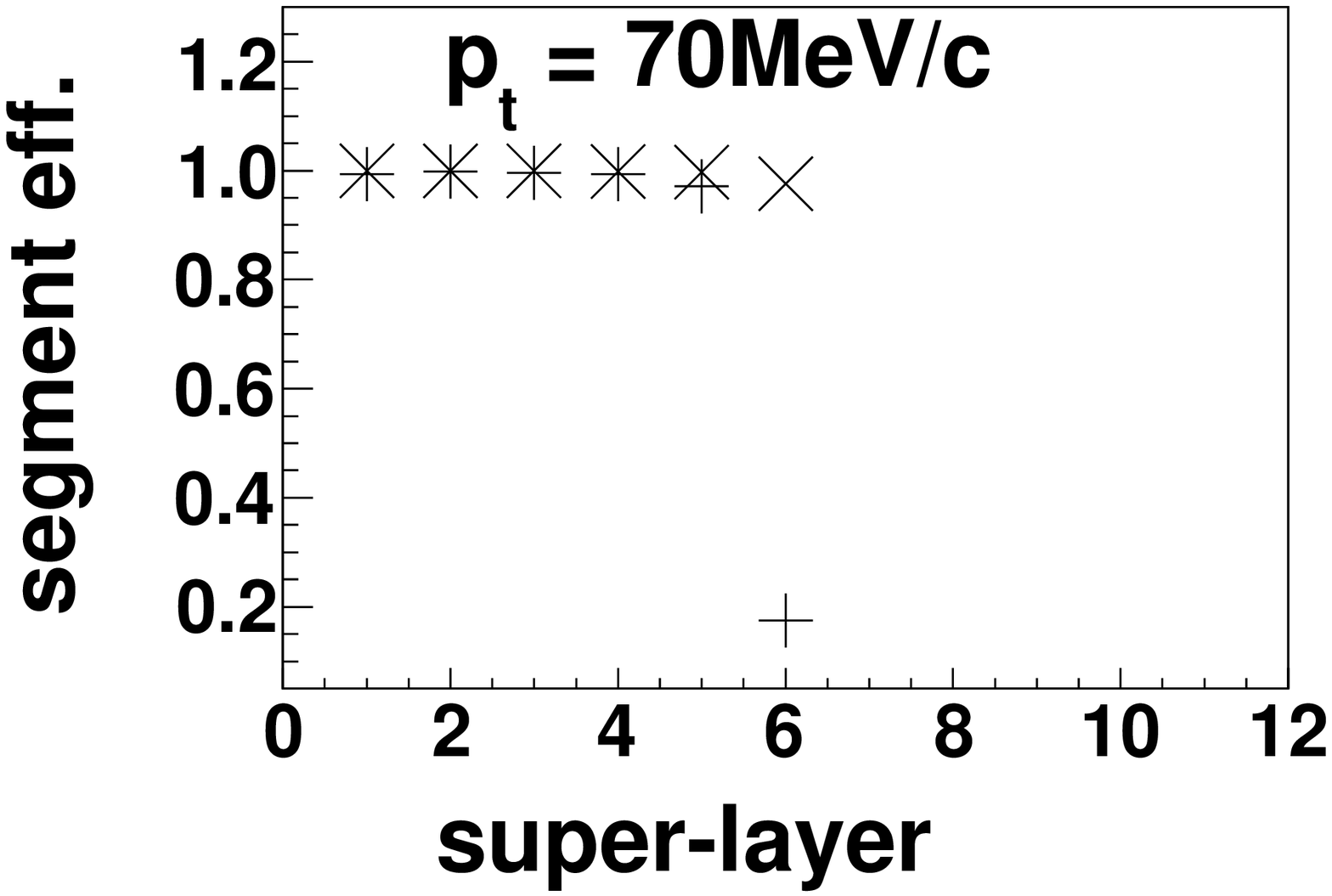}
\includegraphics [height=0.16\textheight,width=0.49\linewidth]
{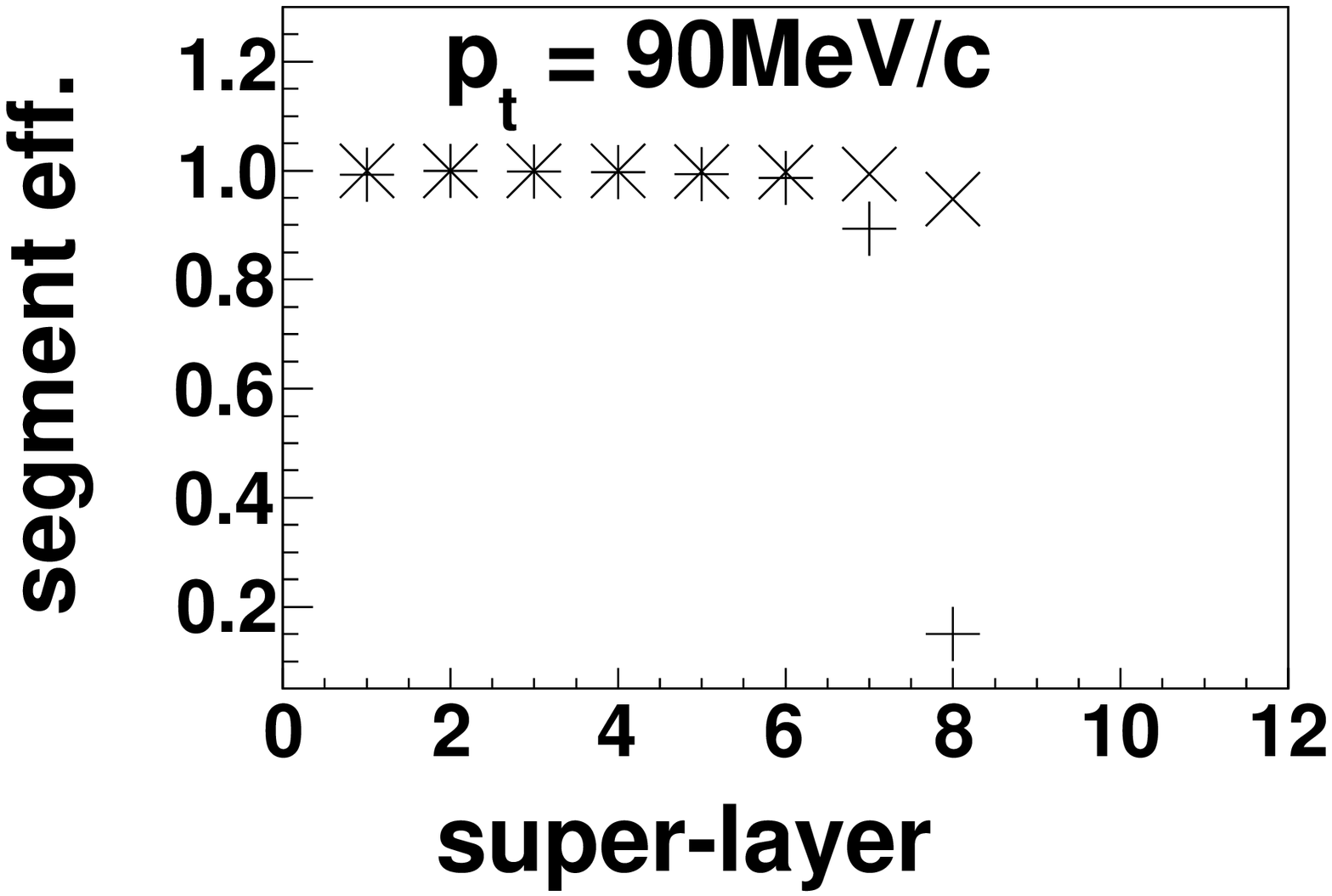}
\includegraphics [height=0.16\textheight,width=0.49\linewidth]
{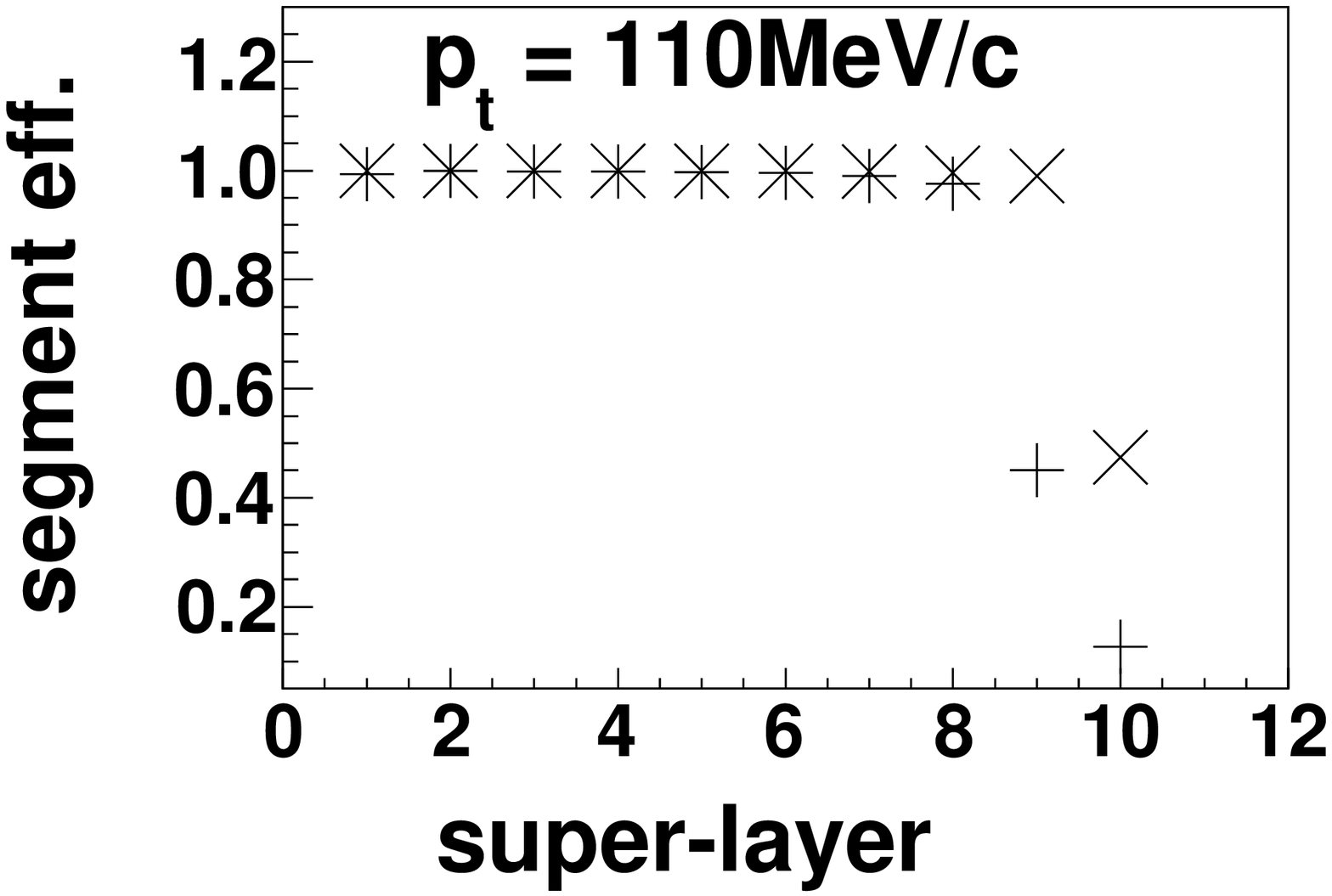}
\renewcommand{\figurename}{Fig.}
\figcaption{
  \label{eff_eff_seg_Psi3770_single_e_neg}
  The segment efficiency of $e^{-}$ with different \Pt~ in each super-layer.
    Symbol $\times$ and $+$ are used to mark the segment efficiencies
    of 14-wire and 8-wire segment pattern dictionaries respectively.
}
\end{center}

\begin{center}
\includegraphics [height=0.24\textheight,width=0.9\linewidth]
{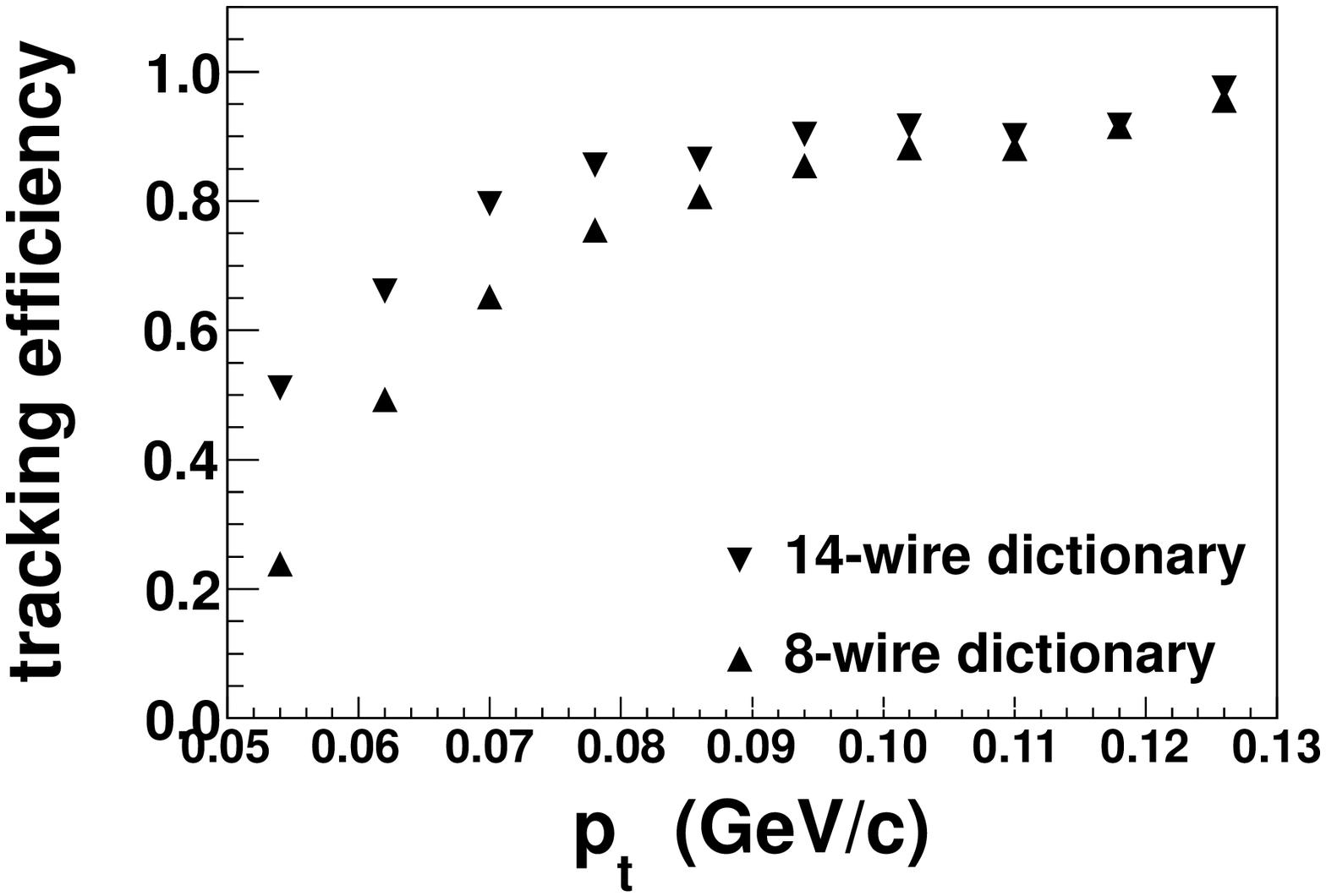}
\renewcommand{\figurename}{Fig.}
\figcaption{
  \label{graph_eff_trk_ZeroVSSix_e_neg_trk_rec_range01}
  The tracking efficiency of $e^{-}$.
}
\end{center}

\subsection{The transverse momentum resolution}
Using the same MC data sample we generate for testing the tracking
efficiency, we compare the transverse momentum resolution
between the extended pattern dictionary and the old pattern
dictionary.
The transverse momentum resolution $\delta{p_{t}}$ is an index of
  track reconstruction quality.  $\delta_{p_{t}}$ is expressed by
Eq. (\ref{eq_del_p})
  \begin{eqnarray}
  \label{eq_del_p}
  \delta{p_{t}}  &=& {p_{t}^\mathrm{rec}} - {p_{t}^\mathrm{MC}},
  \end{eqnarray}
  \noindent where ${p_{t}^\mathrm{rec}} $ is the \Pt~ of a reconstructed
  track, while ${p_{t}^\mathrm{MC}}$ is the \Pt~ of this charged particle
  in MC simulation.  Fig. \ref{hist_del_pt_ZeroVSThree_e_neg} shows the
$\delta{p_{t}}$ distribution of $e^{-}$ (\Pt~ less than 130\,MeV/c)
  reconstructed by the 14-wire and 8-wire segment pattern dictionaries
  respectively. By fit, the $\delta{p_{t}}$ resolution is improved from
  about 0.94\,MeV/c to about 0.89\,MeV/c.

\begin{center}
\includegraphics [height=0.24\textheight,width=0.90\linewidth]
{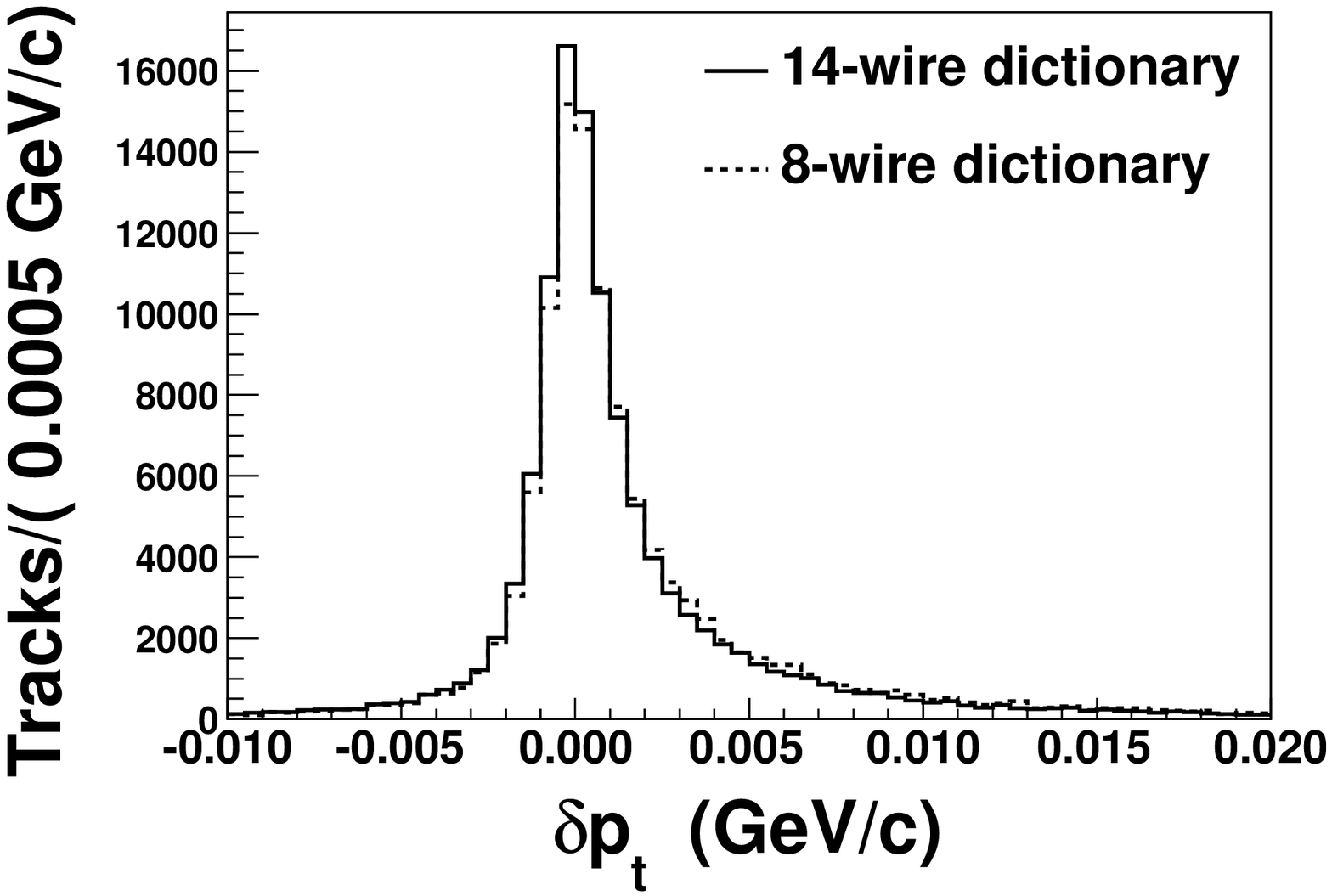}
\renewcommand{\figurename}{Fig.}
\figcaption{
  {\label{hist_del_pt_ZeroVSThree_e_neg}}
  The $\delta{p_{t}}$ distribution of $e^{-}$.
The two histograms are normalized.
}
\end{center}

  \begin{center}
  \includegraphics [height=0.24\textheight,width=1.0\linewidth]
{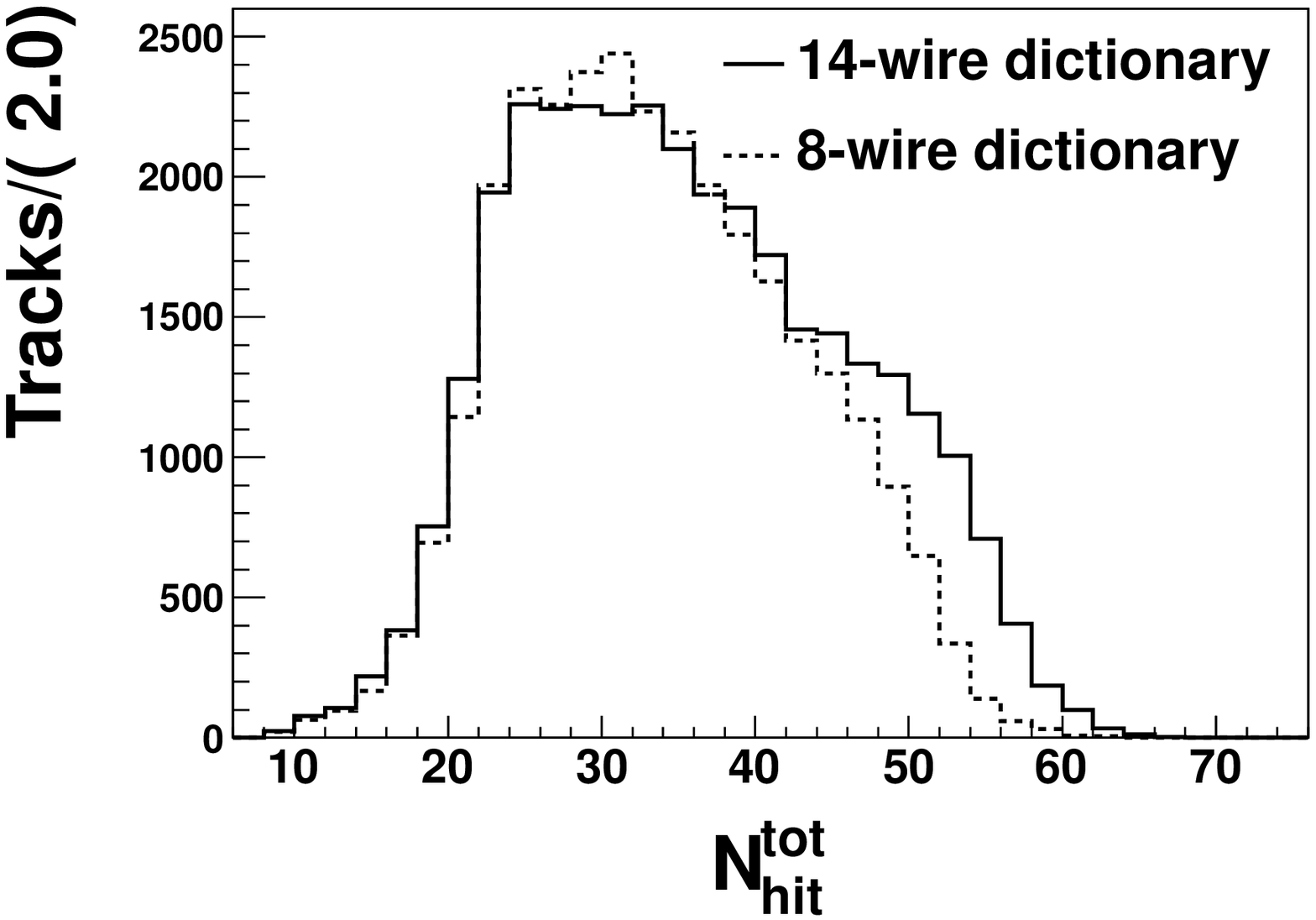}
\renewcommand{\figurename}{Fig.}
\figcaption{
  {\label{hist_num_hits_01_ZeroVSSix_e_pos_trk_rec}}
 The hits number distribution of $e^{-}$.
}
\end{center}

\subsection{The noise resistant ability}
With the same MC events that we use to test the tracking
efficiency, we test the noise resistant ability of the
extended pattern dictionary.
The track hits number is the amount of hits used in the reconstructed
tracks, and it determines the quality of track reconstruction.
The track noise level $R_{\mathrm{noise}}$ can be used to denote the
  anti-noise ability of tracking algorithm. The definition of
$R_{\mathrm{noise}}$  is expressed by Eq.(\ref{eq_noise_level})
  \begin{eqnarray}
  \label{eq_noise_level}
  R_{\mathrm{noise}} &=& \dfrac{N_{\mathrm{hit}}^{\mathrm{noise}}}{N_{\mathrm{hit}}^{\mathrm{tot}}},
  \end{eqnarray}
  \noindent where $N_{\mathrm{hit}}^{\mathrm{noise}}$ is the number
  of fired wires used in track reconstruction but they are not
  fired by the particle; $N_{\mathrm{hit}}^{\mathrm{tot}}$ is the
  number of signal wires used in track reconstruction.
  Figure \ref{hist_num_hits_01_ZeroVSSix_e_pos_trk_rec} and
  \ref{hist_mc_ratio_01_ZeroVSSix_e_pos_trk_rec}  display the number
  of hits on track and $R_{\mathrm{noise}}$  of $e^{-}$ with \Pt~ less
  than 130\,MeV/c. In these plots, the average
  $N_{\mathrm{hit}}^{\mathrm{tot}}$ is increased from 33.1 to 34.9, while
  the average $R_{\mathrm{noise}}$ is decreased from $0.43\%$ to $0.42\%$.
  The change of track noise level is very little.
  \begin{center}
  \includegraphics [height=0.24\textheight,width=1.0\linewidth]
{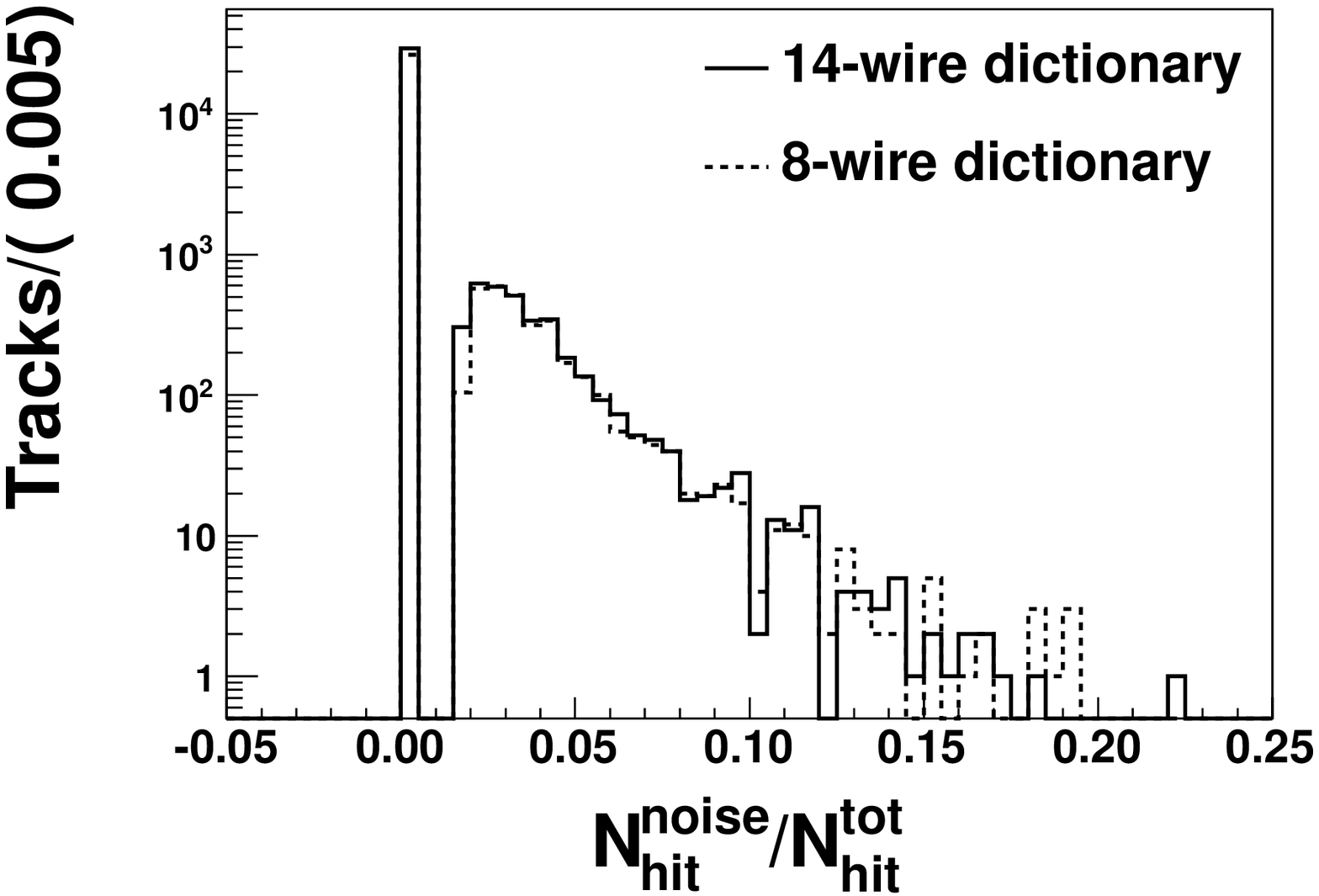}
\renewcommand{\figurename}{Fig.}
\figcaption{
  {\label{hist_mc_ratio_01_ZeroVSSix_e_pos_trk_rec}}
  The $R_{\mathrm{noise}}$ distribution of $e^{-}$.
}
\end{center}

\begin{center}
\includegraphics [height=0.24\textheight,width=1.0\linewidth]
{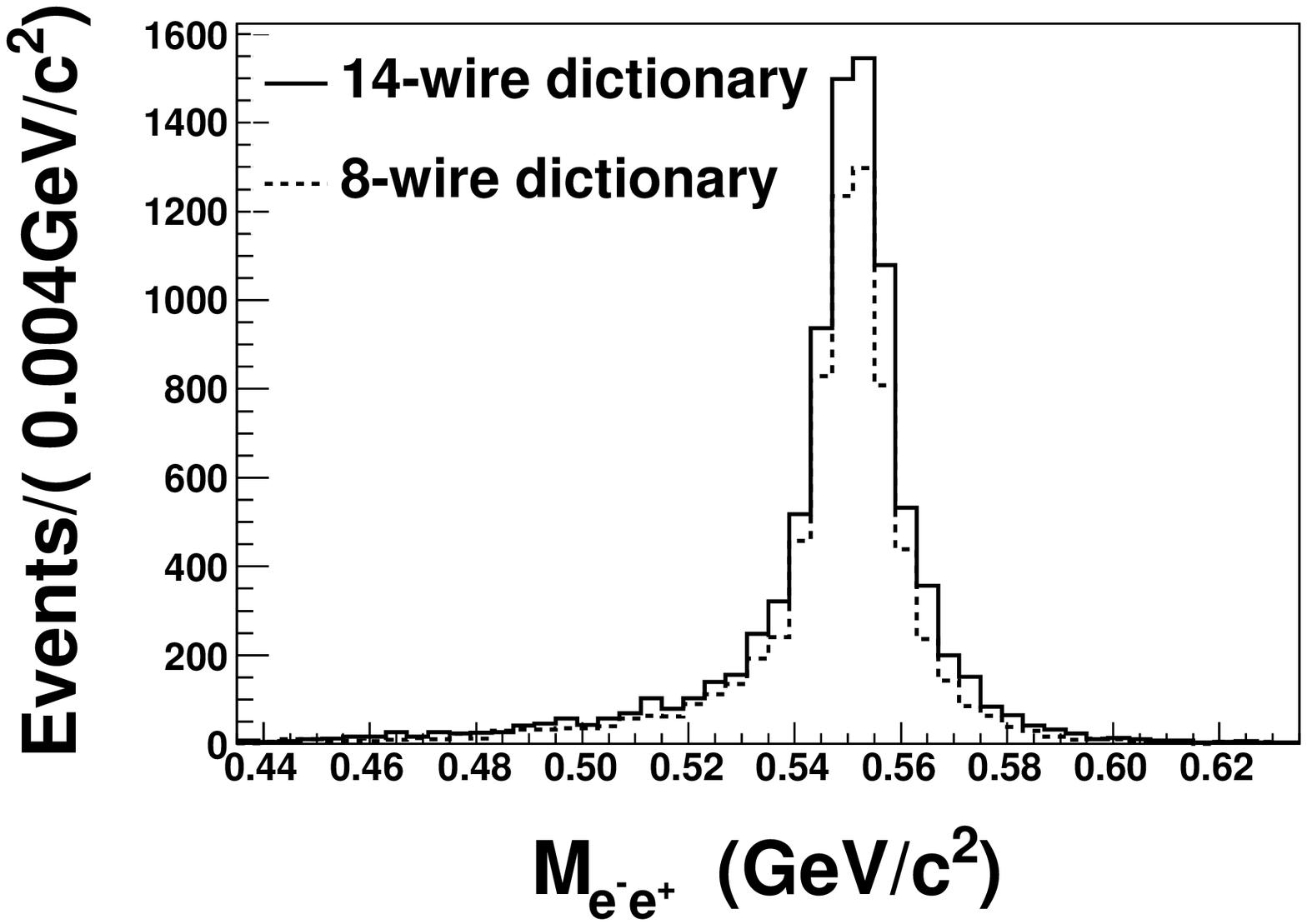}
\renewcommand{\figurename}{Fig.}
\figcaption{
  {\label{hist_mraw_Pp_1_PpGChicj2KEta2Ex4cf}}
  The reconstructed signals of $\eta$ from
    MC decay sample $\psi(2S)\to\gamma\chi_{CJ}\to\gamma \eta K^{+} K^{-}$ with
    the requirement \Pt~ of $e^{-}$ less than 130\,MeV/c.
}
\end{center}

\subsection{Application in physics analysis}
We generate a physics process, $\psi(2S) %
\to\gamma\chi_{cJ} \to\gamma K^{+} K^{-}\eta \to \gamma  K^{+} K^{-} e^{+}
e^{-}$, to test the performance of the new segment pattern dictionary.
Figure \ref{hist_mraw_Pp_1_PpGChicj2KEta2Ex4cf} is the invariant mass
distribution of $\eta$. As shown in the plot, the new segment pattern
dictionary can keep more physical events than the old segment pattern
dictionary. The total reconstruction efficiency is increased by about
$4\%$; if we only count the events with \Pt~ of $e^{-}$ less than
130\,MeV/c, the efficiency is increased by about $20\%$.

\subsection{Application in experimental data}
We test the new segment pattern dictionary with experimental data.
Figure \ref{hist_pt_mdc_trk_1_mdc_trk} shows the number of reconstructed
tracks with respect to \Pt~ for a sample of $\psi(3770)$ experimental data.
The amount of reconstructed tracks with \Pt~ less than 130\,MeV/c is
increased by about $15\%$.
The number of charged tracks of a $\psi(3770)$ event is usually not less
than 4. The differences in the number of charged tracks
and in the track angular distribution
between
$\psi(3770)$ events and single $e^{-}$ MC events
make the difference between the curves of reconstructed tracks in
Fig. \ref{hist_pt_mdc_trk_1_mdc_trk}
and efficiency curves in
Fig. \ref{graph_eff_trk_ZeroVSSix_e_neg_trk_rec_range01}.
\begin{center}
\includegraphics [height=0.24\textheight,width=0.9\linewidth]
{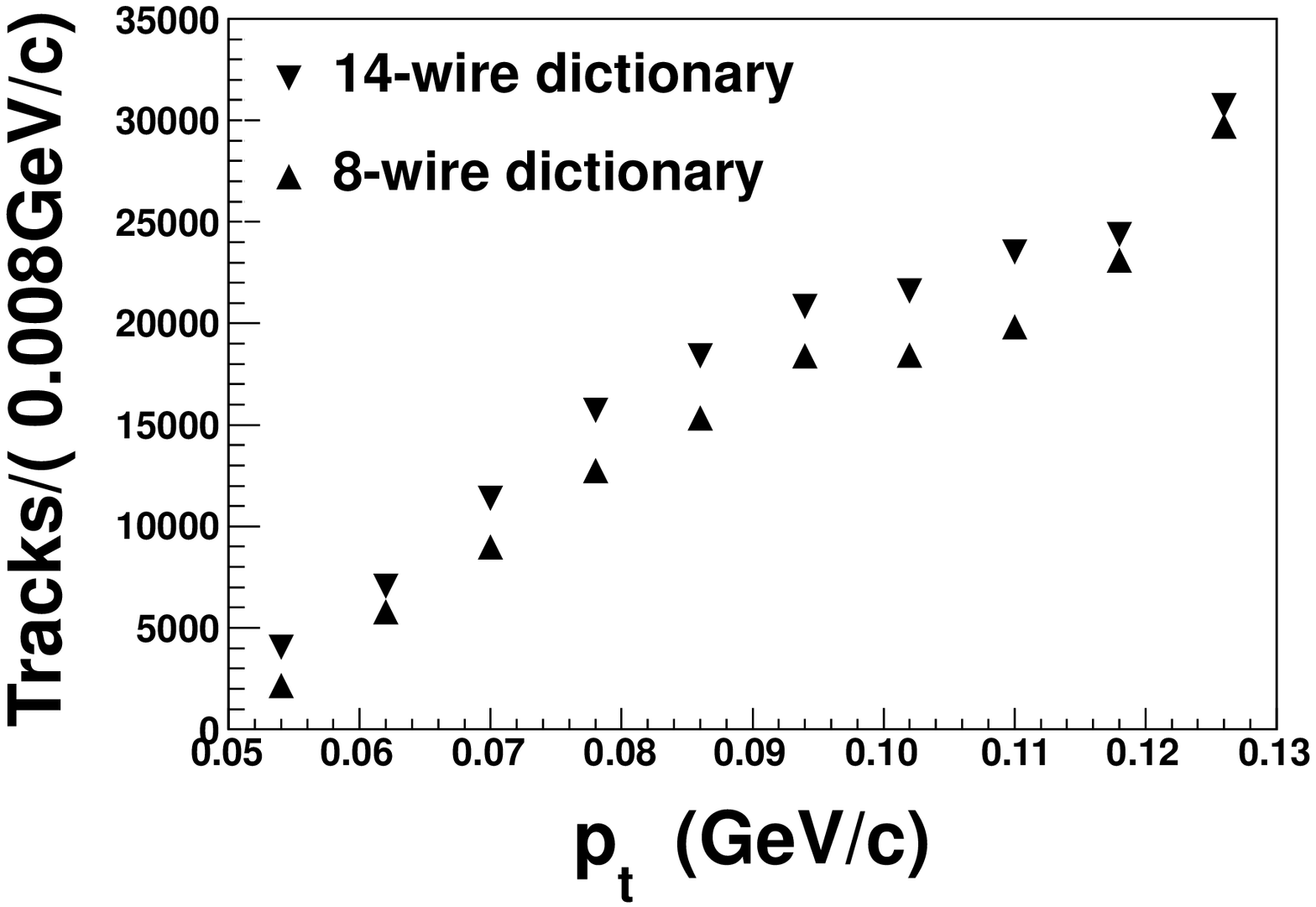}
\renewcommand{\figurename}{Fig.}
\figcaption{
  \label{hist_pt_mdc_trk_1_mdc_trk}
  The reconstructed tracks numbers from $\psi(3770)$ experimental data.
}
\end{center}

\subsection{CPU-time consumption}
Within BESIII offline software system~\cite{boss}, We test
the CPU-time increase of MdcPatRec using the 14-wire
segment pattern dictionary. Compared with using the 8-wire
segment pattern dictionary, the CPU-time of MdcPatRec
in segment finder is increased by about 2\,ms for an event on the average; the
CPU-time in total MdcPatRec algorithm is increased by about $30\%$;
the CPU-time of the total event reconstruction is increased by about $10\%$.

\section{Summary and discussion}
In this paper, we study the issue of low tracking efficiency of
the MdcPatRec algorithm in the low $p_t$ range and understand one
source of the problem as the shortage of 8-wire segment group: its
limit of azimuth angle coverage and wire group asymmetry. We propose
a 14-wire group and its segment pattern dictionary as a remedy to
the defect. The performance shows that the new wire group
construction scheme and its segment pattern dictionary can improve
the tracking efficiency for the low \Pt~  tracks. However, the
CPU-time consumption is increased, so we will do more work to optimize
the algorithm.

\end{multicols}

\vspace{-1mm}
\centerline{\rule{80mm}{0.1pt}}
\vspace{2mm}

\begin{multicols}{2}

								 \end{multicols}

								 \clearpage

								 \end{document}